\documentclass[useAMS,usenatbib]{mn2e}
\usepackage{epsfig}
\usepackage{graphicx}
\usepackage{deluxetable}
\usepackage{natbib}

\newcommand{\fermi}{\textit{Fermi~}}
\newcommand{\swift}{\textit{Swift~}}

\title[\textit{Fermi}/GBM-\textit{Swift}/BAT joint spectral analysis]{Spectral and temporal analysis of the joint \textit{Swift}/BAT-\textit{Fermi}/GBM GRB sample}
\author[F. J. Virgili, Y. Qin, B. Zhang and E.-W. Liang]{Francisco J. Virgili$^{1,2}$\thanks{E-mail:
fjv@astro.livjm.ac.uk (FJV)}, Ying Qin$^{3}$, Bing Zhang$^{2}$, and Enwei Liang$^{3}$\\ 
$^{1}$Astrophysics Research Institute, Liverpool John Moores University, Birkenhead CH41 1LD, UK\\
$^{2}$Department of Physics and Astronomy, University of Nevada Las Vegas, Las Vegas, NV 89154, USA \\
$^{3}$ Department of Physics and Astronomy, Guangxi University, Nanning 530004, China
}

\begin{document}

\date{Accepted xxxx. Received xxxxx; in original form xxxxx}

\pagerange{\pageref{firstpage}--\pageref{lastpage}} \pubyear{2011}

\maketitle

\label{firstpage}

\begin{abstract}

Using the gamma-ray bursts simultaneously detected by \textit{Swift}/BAT and \textit{Fermi}/GBM we performed a joint spectral and temporal analysis of the prompt emission data and confirm the rough correlation between the BAT-band photon index $\Gamma^{\rm BAT}$ and the peak spectral energy $E_{\rm peak}$. With the redshift known sub-sample, we derived the isotropic gamma-ray energy $E_{\rm \gamma,iso}$ and also confirm the $E_{\rm \gamma,iso}-E_{\rm peak,rest}$ relation, with a larger scatter than the Amati sample but consistent with GBM team analyses. We also compare the $T_{90}$ values derived in the GBM band with those derived in the BAT band and find that for long GRBs the BAT $T_{90}$ is usually longer than the GBM $T_{90}$, while for short GRBs the trend reverses. This is consistent with the soft/hard nature of long/short GRBs and suggests the importance of an energy-dependent temporal analysis of GRBs. 

\end{abstract}


\section{Introduction}

Since the launch of the \textit{Fermi} satellite in 2008, a significant number of gamma-ray bursts (GRBs) have been observed by both the \fermi Gamma-ray Burst Monitor (GBM, 8 keV-40 MeV) and \swift \cite{gehrels04}. Since the energy band of \textit{Swift} BAT is narrow (15-150 keV) and the peak of the spectrum is often outside the observed energy window \cite{sakamoto09}, the joint BAT-GBM sample is valuable for \textit{Swift} science by providing important information about the prompt emission, especially the characteristic peak energy of the $\nu F_{\nu}$ spectrum, $E_{\rm peak}$. Using the \textit{Fermi}/GBM-\textit{Swift}/BAT joint sample we performed a two-part analysis on the prompt emission data.  First, a joint spectral analysis where we fit the data from both detectors and investigated several empirical correlations.  Second,  a temporal analysis where we rigorously derived $T_{90}$ in the 8-1000 keV band and compared it to the published BAT values.  

A variety of empirical relations have been discussed in the literature \citep{amati02,amati03,sakamoto04,sakamoto06,ghirlanda04,liangzhang05,yonetoku04}, many of which utilize $E_{\rm peak}$. Since $E_{\rm peak}$ cannot be well measured for most \swift GRBs, efforts have previously been made to look for indicators of $E_{\rm peak}$ based on the available observed quantities. In particular, the effective photon power law index in the BAT band, $\Gamma^{\rm BAT}$, has been found to be broadly correlated with $E_{\rm peak}$. Two versions of this relation are found in the literature: one presented in Sakamoto et al. (2009) using simulations of BAT spectra and $E_{\rm peak}$ measurements from broad-band detectors (e.g. Konus-Wind), and another presented in Zhang et al. (2007b) with $E_{\rm peak}$ estimates based on the hardness ratio in the BAT band itself (Zhang et al. 2007a; see also Cui et al. 2005). In \S3.2 we re-calibrate this correlation with the GBM-BAT joint sample. The $E_{\rm \gamma,iso} - E_{\rm peak,rest}$ relation \citep[or Amati relation;][]{amati02,amati03} is one of the most widely studied, and hotly debated, empirical correlations that may connect to GRB physics  \citep[e.g.][]{band05,nakar05,kocevski11}. In \S3.3 we test the Amati relation with the redshift-known sub-sample of the GBM-BAT joint sample.

Finally, traditionally long and short GRBs were defined in the BATSE band with a separation of about 2 seconds in the observed frame (Kouveliotou et al. 1993).  When applied to \swift GRBs and combined with multi-wavelength afterglow data,  this definition creates confusion when used for GRB classification (e.g. Zhang et al. 2007b, 2009). Since the energy band of GBM is similar to that of BATSE, a comparison between the measured $T_{90}$ durations in the two detectors is of great interest and is performed in \S4. 

\section{Sample and Data Reduction}

We conducted our analysis on the sample of GRBs that were observed simultaneously with the \textit{Swift}/BAT and \textit{Fermi}/GBM detectors.  The sample was chosen from bursts that were consistent both temporally (similar trigger times in \swift MET) and in sky placement (RA/DEC) between June 2008 and May 2011.  Our final sample consists of 75 bursts whose spectra were fit individually to check consistency with the literature and then fit jointly.  Some bursts were removed because they were a \swift ground detection (e.g. 100427A; partial data) or a false detection (e.g. 080822B which has been classified as an SGR flare). 

The \textit{Fermi}-GBM GRB data was downloaded from the online \fermi GRB burst catalog found on the NASA HEASARC website\footnote{http://heasarc.gsfc.nasa.gov/W3Browse/fermi/fermigbrst.html} and the time-integrated spectra were fit with the GBM software \verb"RMfit"\footnote{fermi.gsfc.nasa.gov/ssc/data/analysis/}. The \textit{Swift}-BAT data was downloaded from the \swift data table, also located on the NASA HEASARC website\footnote{http://heasarc.gsfc.nasa.gov/cgi-bin/W3Browse/swift.pl}.  The resulting data was processed using the typical \textit{Swift}-BAT products (\verb"batgrbproduct") after updating to the current calibration database (\verb"bateconvert") to create the burst lightcurves and spectra.

\section{Spectral analysis}

\subsection{Results}

We performed spectral fits for the GBM GRBs with \verb"RMfit" using the built-in power law (PL), power law with exponential cutoff (CPL, 100 keV normalization energy) and Band function \cite{band93} models over the entire burst temporal and spectral (8 keV-40 MeV) interval in order to determine the best fit model. The goodness of the fits are measured with the C-statistic. If the C-stat was comparable for two models, the simpler model (i.e. the one with less parameters) was chosen as the better fit to compensate for the added degree of freedom.  On many occasions, the decrease in the statistic was greater than 30-50 for a PL versus a CPL fit, for example, but not significantly reduced when comparing the CPL and Band function fit.  Evidence of this can be seen in the majority of CPL fits in the GBM data summarized in Tables 1 and 2. The spectral regime chosen for the NaI detectors was roughly 8-900 keV and 200 keV-40 MeV for the BGO detectors.  Our spectral fit results are generally consistent with the literature (e.g. Bissaldi et al. 2011; Nava et al. 2011; individual burst GCN reports).

For BAT spectral fitting, it is important to make sure that the the satellite was not slewing during data acquisition.  If \swift was slewing while capturing a spectrum, a new weighted response file was created to compensate for the motion \cite{sakamoto11}.  The response matrices are created in 5 second intervals and weighted to created a final response file that is used with the total spectrum.  Each time-integrated spectrum was fit over the full BAT energy range of 15-150 keV using the same three models listed above.  If a model showed improvement greater than $\Delta \chi^2 >$ 6 per added degree of freedom then that model was deemed a better fit. Our spectral fitting for the BAT data and joint spectral fits for the GBM and BAT observed spectra are performed with Xspec. We re-bin the spectra with the criterion that the number of photon counts per bin is greater than 20 by using the HEAsoft tool {\em grppha} if the burst was sufficiently short or weak that this procedure was warranted.  In order to address the calibration between the BAT and the GBM NaI/BGO detectors, each spectrum was fit with a varying calibration constant. The BAT calibration was fixed to 1.0 and the NaI and BGO constants were left as free parameters to be fit.  About half of the bursts did not produce reliable fits to the BGO detector calibration constant, likely due to the sometimes low signal in the BGO detector.  For these bursts the BGO constant was fixed to the mean value of the distribution of BGO calibration constants (1.3) and the spectra were refit.  The $\chi^2$ and spectral parameter errors were generally reduced when compared to the fits where all the calibration constants were forced to be equal.

\begin{figure*}
\centerline{\includegraphics[width=9.2cm]{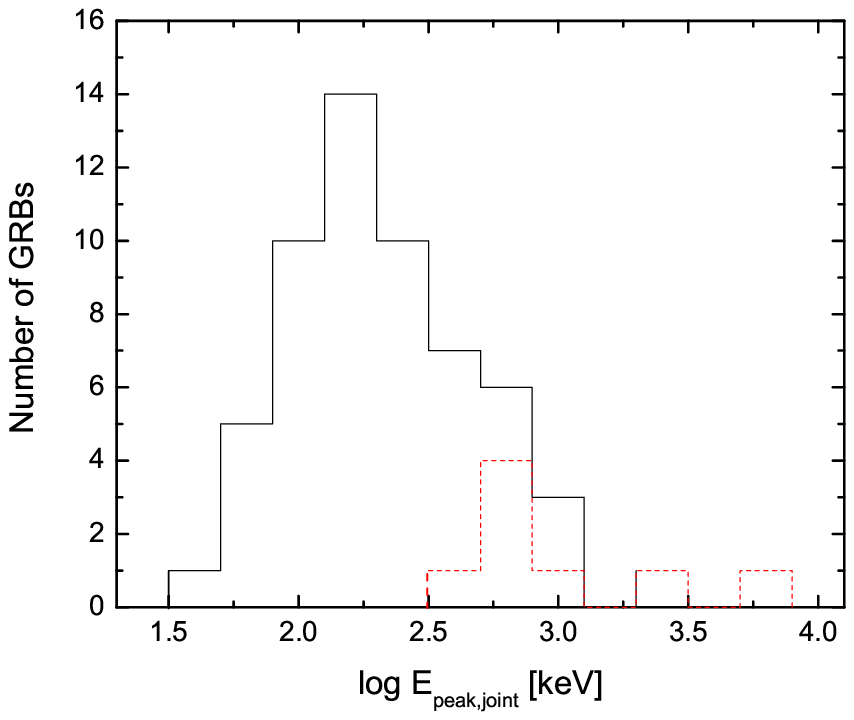}\includegraphics[width=9.2cm]{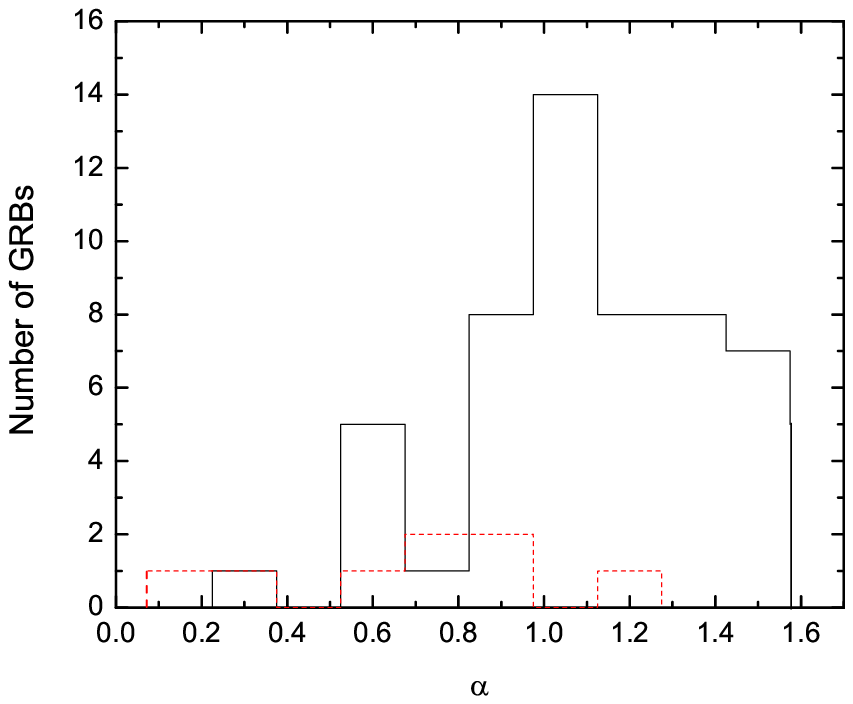}}
\caption{Distributions of $E_{\rm peak}$ and $\alpha$ for the best fit joint sample bursts.  Both are generally consistent with the literature \citep[e.g.][]{preece00,zhang11,nava11}.  The solid (black) and dashed (red) histograms are for long and short GRBs, respectively.}
\label{distributions}
\end{figure*}

The best-fit spectral models of all the 75 bursts in the joint GBM-BAT sample are presented in Table 1 (simple powerlaw model) and Table 2 (cutoff powerlaw or Band models). The distributions of $E_{\rm peak}$ and $\alpha$ of the sample are shown in Fig. \ref{distributions}. The results are generally consistent with previous results \citep[e.g.][]{preece00,zhang11,nava11}. The $E_{\rm peak}$ distribution is roughly log-normal but is somewhat wider than that of the bright BATSE GRB sample (Preece et al. 2000).  Our sample includes GRBs in a broader $E_{\rm iso}$ range, broadening the observed $E_{\rm peak}$ range. The low-energy spectral index $\alpha$ peaks around 1 as suggested by previous works. We do not present the distribution of the high energy photon index $\beta$ due to the small sample size (most spectra can be fit by a cutoff powerlaw model). Although they form a limited part of our sample the distribution for short bursts show behavior consistent with previous samples, namely a harder median E$_{\rm peak}$ and similar or slightly shallower pre-break slope.

\begin{figure*}
\centerline{\includegraphics[width=15cm]{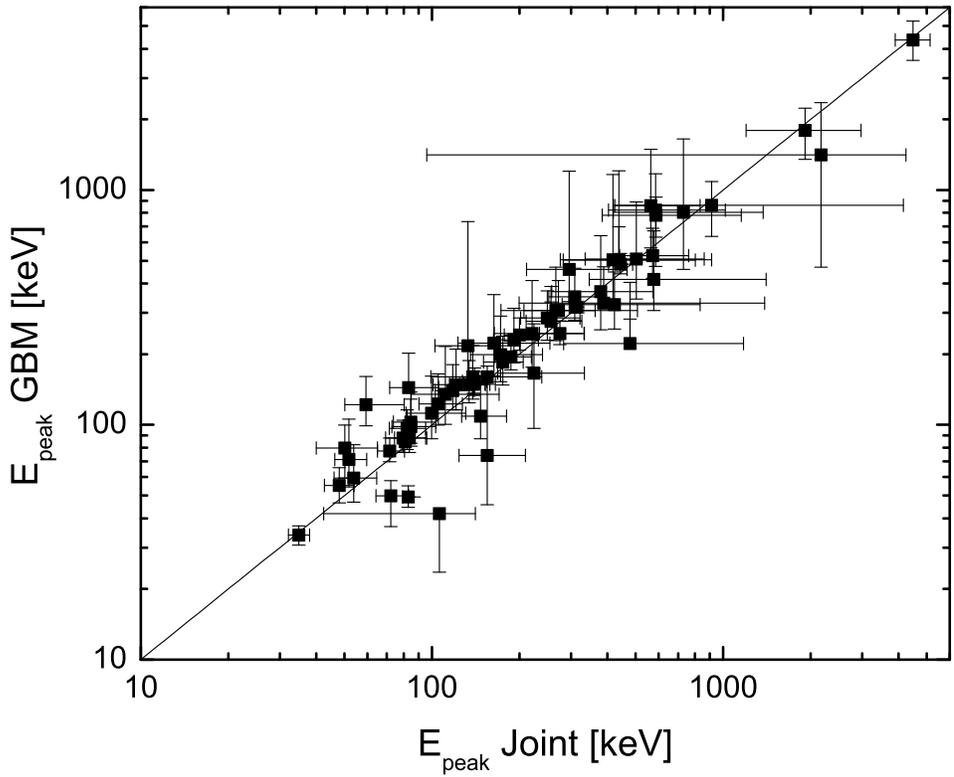}}
\caption{The relationship between the best fit $E_{\rm peak}$ for the joint sample and the GBM spectra.  The line superimposed is the line with slope 1.  This shows that adding the BAT data into the fit does not significantly change the values of $E_{\rm peak}$.}
\label{epep2}
\end{figure*}

Since the GBM energy band covers that of BAT, we also derive $E_{\rm peak}$ with the GBM data only and compare the values with the joint fit values. Figure \ref{epep2} shows the relationship between the separately derived GBM and joint sample $E_{\rm peak}$ values with a line of slope 1 drawn for reference, showing the generally small differences between the samples.   

\subsection{$\Gamma^{\rm BAT}-E_{\rm p}$ relation}

With our spectral analysis results, we examine the empirical relations between the PL slope of the BAT spectrum, $\Gamma^{\rm BAT}$, and $E_{\rm peak}$ proposed by Sakamoto et al. (2009) and Zhang et al. (2007b):
\begin{equation}
\log E_{\rm peak}~=~3.258~-~0.829\Gamma^{\rm BAT}
\end{equation}
and 
\begin{equation}
\log E_{\rm peak}~=~(2.76~\pm~0.07)~-~(3.61~\pm~0.26)~ \log \Gamma^{\rm BAT},
\end{equation}
respectively\footnote{Sakamoto et al. (2009) derive 1$\sigma$ error levels which are based on the value of $\Gamma$ (see their Table 1): Lower limit: -20.684 + 43.646$\Gamma$ - 26.891$\Gamma^2$ + 5.185$\Gamma^3$ Upper limit: -5.198 + 16.568$\Gamma$ - 10.630$\Gamma^2$ + 2.034$\Gamma^3$}.  Both Zhang et al. (2007a,b) and Sakamoto et al. (2009) indicate that the relation is valid between slopes of roughly $1.2 < \Gamma^{\rm BAT} < 2.3$.  Here we aim to present the entire data set of the joint sample and see how the relation fares over the entire observable sample.  It is also interesting to explore whether using the joint spectra is consistent with the derivation of $E_{\rm peak}$ from using only the GBM sample and a motivation for investigating the combined constraints of the two detectors.  

\begin{figure*}
\centerline{\includegraphics[width=9.2cm]{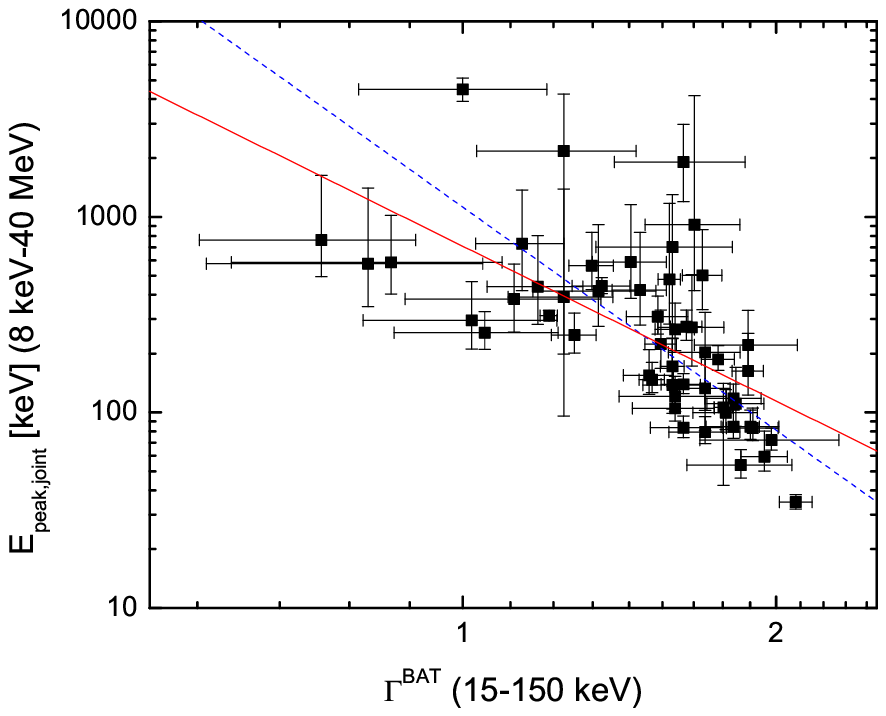}\includegraphics[width=9.2cm]{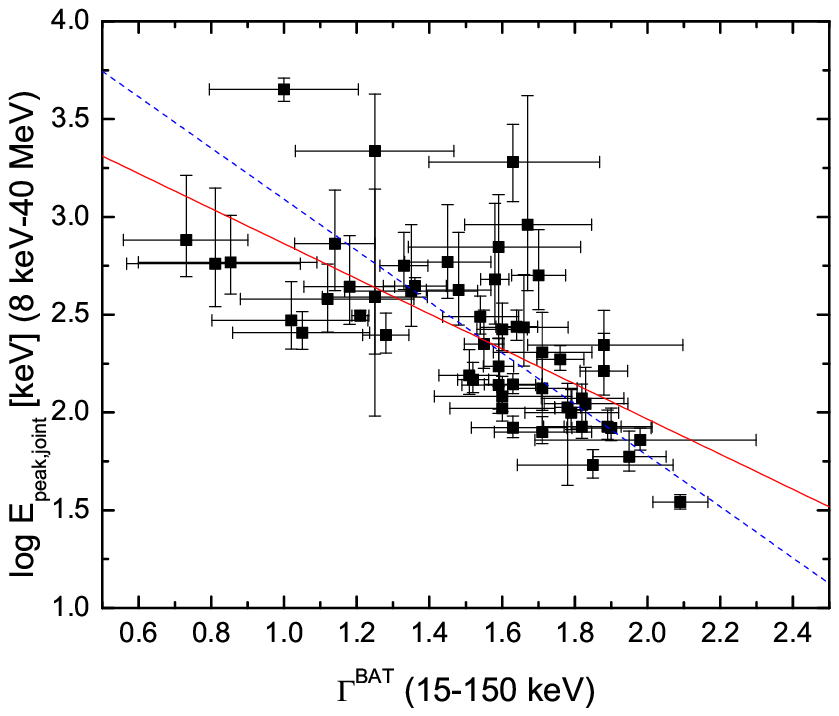}}
\caption{The $\Gamma^{\rm BAT}-E_{\rm peak}$ relation in both log-log (a) and log-linear (b) space created from 53 best-fit spectra.  15 bursts were removed from the original sample of 75 because the BAT best-fit model was a CPL and not a simple PL. An additional seven bursts were removed because the joint spectrum is described by a simple PL.  As shown below, the difference in $E_{\rm peak}$ between the CPL and Band model fit for the joint spectra is negligible, so the best-fit model value was used.  The lines superimposed on the plots are the best-fit lines generated using an ordinary least squares fit (red, solid) and a least squared linear bisector fit (blue, dash) that considers some of the scatter in the relation.  The latter procedure's parameters are summarized in the text.}
\label{best_fit}
\end{figure*}

Using only the 53 joint sample bursts that had BAT spectra best fit with a PL function and the joint spectra $E_{\rm peak}$, regardless of whether it is from a CPL or Band function, we derive relations that are steeper than the values above (Figure \ref{best_fit}).  The ordinary least squares method gives shallower slopes more consistent with the values of Sakamoto (2009) and the bisector fits are steeper and more consistent with the values presented in Zhang et al. (2007b).  Below we summarize the bisector fits, assuming these produce more reliable fits that take into account the intrinsic scatter of the distribution:

\begin{equation}
\log E_{\rm peak}~=~(4.40~\pm~0.509)~-~(1.31~\pm~0.148)\Gamma^{\rm BAT}
\end{equation}
\begin{equation}
\log E_{\rm peak}~=~(3.05~\pm~0.359)~-~(3.79~\pm~0.554)~ \log \Gamma^{\rm BAT}
\end{equation}

\begin{figure*}
\centerline{\includegraphics[width=15cm]{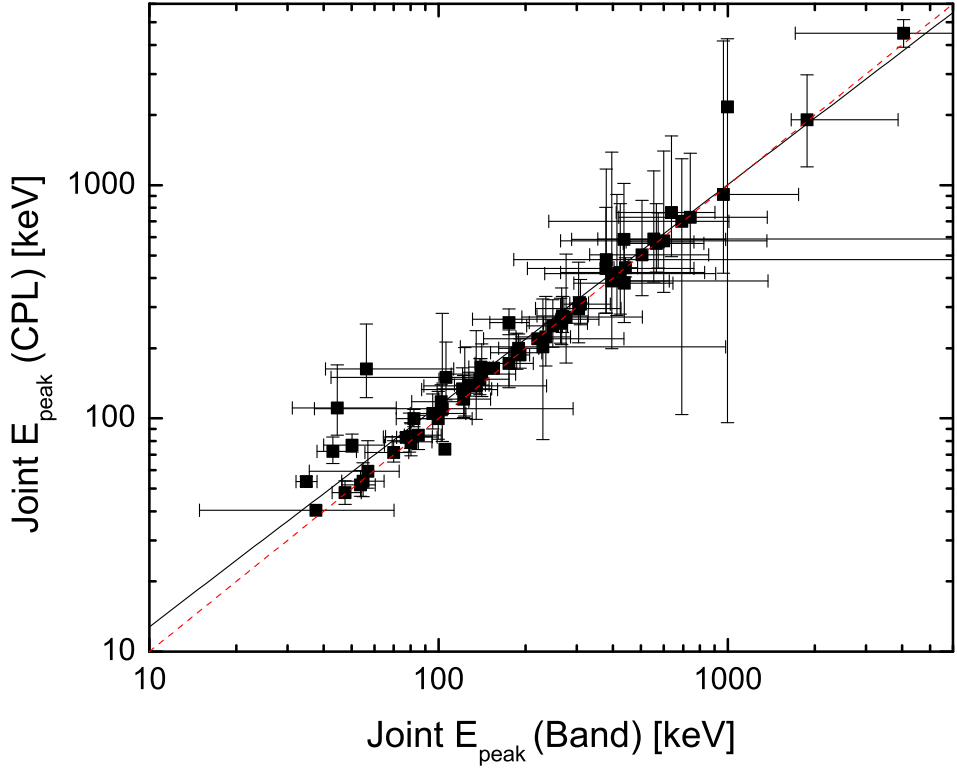}}
\caption{The relationship between the values joint $E_{\rm peak}$ for a CPL and Band model for all 75 GRBs in our sample.  The best fit line (solid, black) has a slope of nearly 1, as shown by the reference line (dashed, red), with very few outliers from this relation.  Such small changes in the value are negligible in the global analysis of spectral properties and we, therefore, always take the best-fit model value of $E_{\rm peak}$.  Some Band function fits are not well constrained and are reflected in the large horizontal error bars for some bursts.}
\label{epep1}
\end{figure*}

Using the values of $E_{\rm peak}$ from either a CPL or Band function fit is a valid assumption, as shown in Figure \ref{epep1}.  This figure shows the values of $E_{\rm peak}$ derived from both models together with the best-fit line, which has a slope very close to 1.

\begin{figure*}
\centerline{\includegraphics[width=9.2cm]{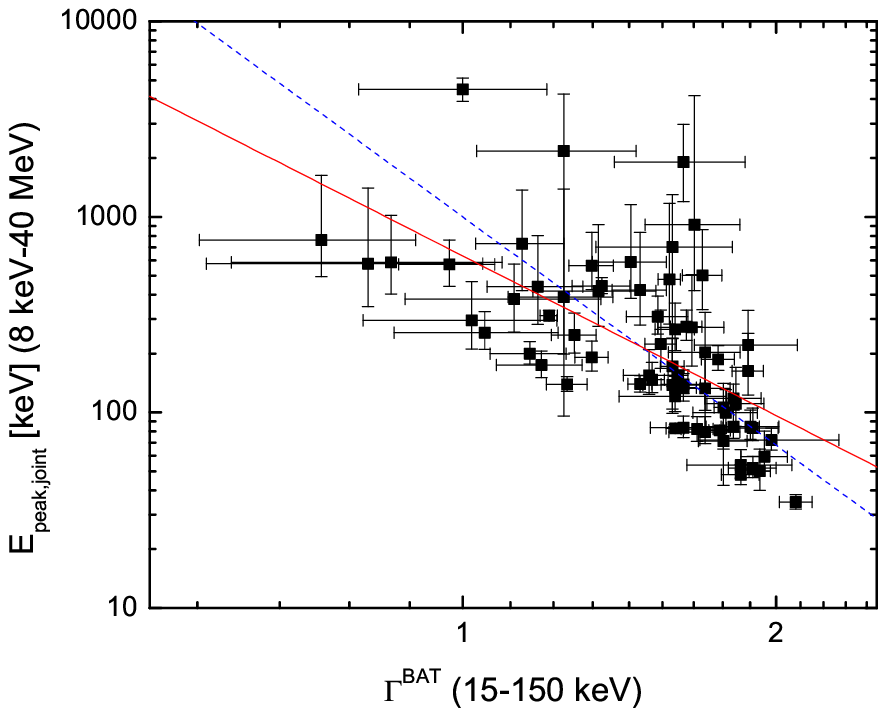}\includegraphics[width=9.2cm]{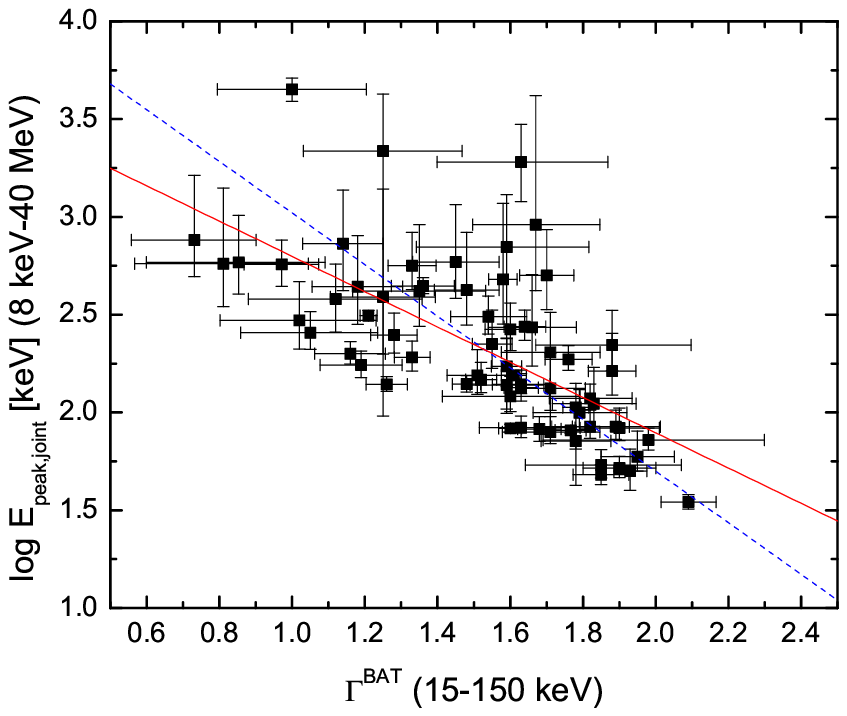}}
\caption{The $\Gamma^{\rm BAT}-E_{\rm peak}$ relation in both log-log (a) and log-linear (b) space created from all the bursts in our sample.  We refit the \swift bursts that are described as a CPL with a simple PL model and use this value of $\Gamma$, regardless of the increase in the $\chi^2$.  The lines superimposed on the plots are the best-fit lines generated using an ordinary least squares fit (red, solid) and a least squared linear bisector fit (blue, dash) that considers some of the scatter in the relation.  The latter procedure's parameters are summarized in the text.}
\label{PL_fit}
\end{figure*}

Next we look at all the bursts in the sample, including those whose BAT spectra are fit with a CPL model.  Since the relations above deal with the PL slope of the BAT spectrum, we forced the remaining CPL bursts to be fit with a PL model to see if the relationships changed in any way (Fig \ref{PL_fit}).  The results are similar, with the slopes becoming slightly steeper than the previous values:

\begin{equation}
\log E_{\rm peak}~=~(4.34~\pm~0.475)~-~(1.32~\pm~0.129)\Gamma^{\rm BAT}
\end{equation}
\begin{equation}
\log E_{\rm peak}~=~(3.00~\pm~0.332)~-~(3.88~\pm~0.469)~ \log \Gamma^{\rm BAT}
\end{equation}

\begin{figure*}
\centerline{\includegraphics[width=15cm]{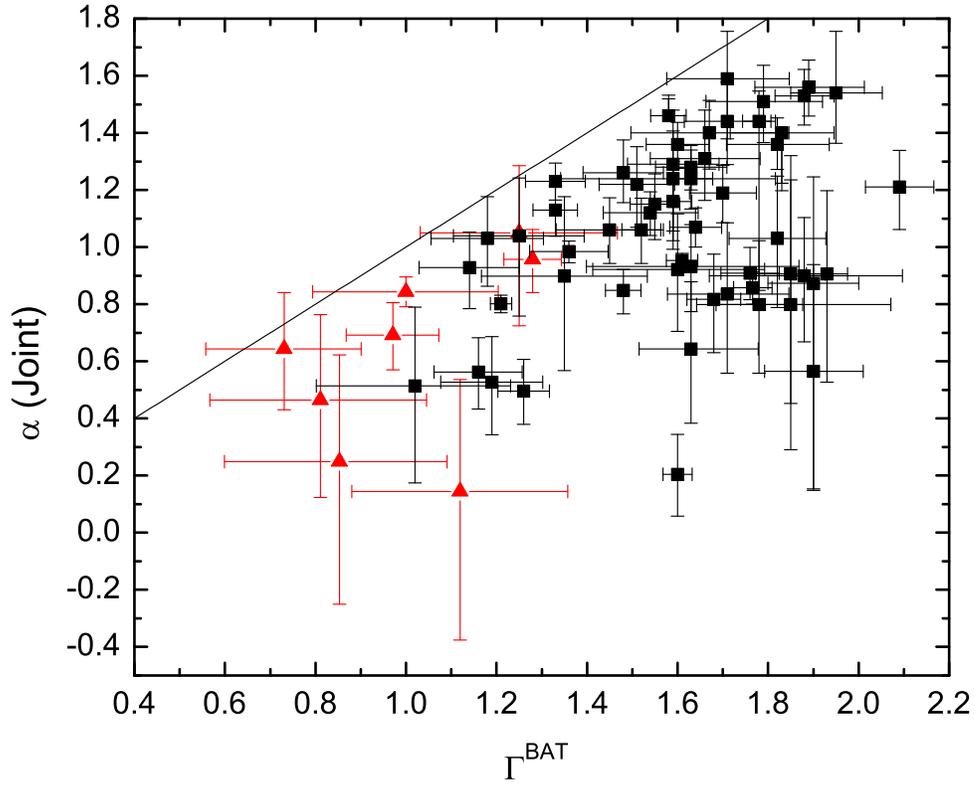}}
\caption{Relationship between the BAT PL spectral slope $\Gamma^{\rm BAT}$ and the joint fit CPL/Band $\alpha$ for short (red, triangle) and long (black, square) bursts.  The line superimposed is a reference of slope = 1.}
\label{ga}
\end{figure*}

Lastly, Figure \ref{ga} shows the relationship between $\Gamma^{\rm BAT}$ and the joint CPL/Band fit spectral slope $\alpha$.  The data all lie below the reference line of $\alpha=\Gamma^{\rm BAT}$. This is understandable, since the effective photon index $\Gamma^{\rm BAT}$ has to compensate for the curved spectrum around $E_{\rm peak}$ and is, therefore, generally steeper. The scatter in the relative steepness is related to the location of $E_{\rm peak}$. The larger the $E_{\rm peak}$, the closer $\Gamma^{\rm BAT}$ to $\alpha$. If $E_{\rm peak}$ is relatively low, the BAT-band spectra are severely bent and the effective $\Gamma_{\rm BAT}$ could be very different from $\alpha$.

\subsection{Amati Relation}

\begin{figure*}
\centerline{\includegraphics[width=9.2cm]{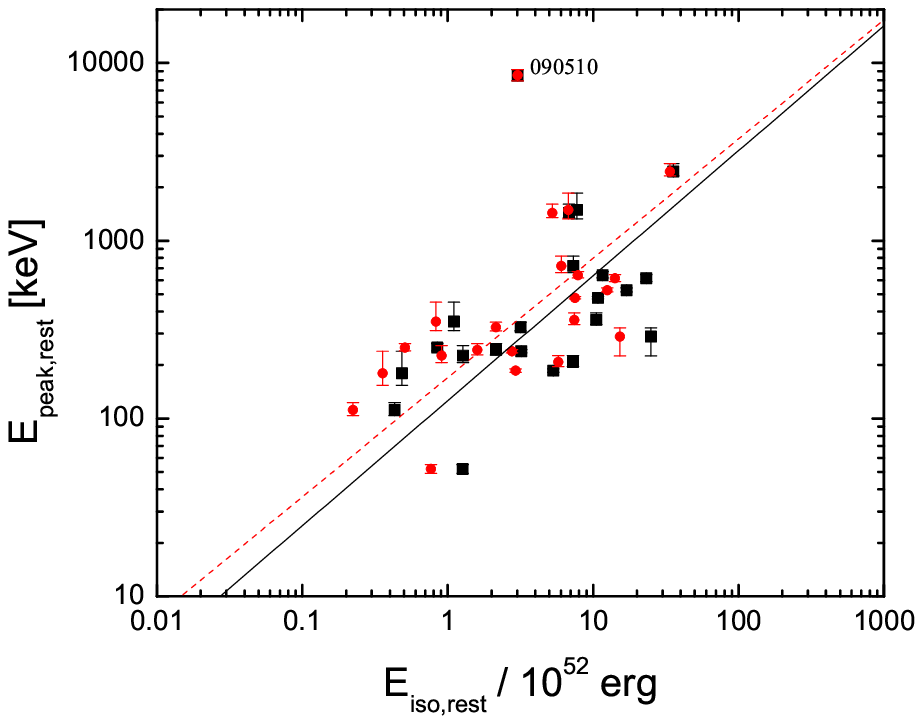}\includegraphics[width=9.2cm]{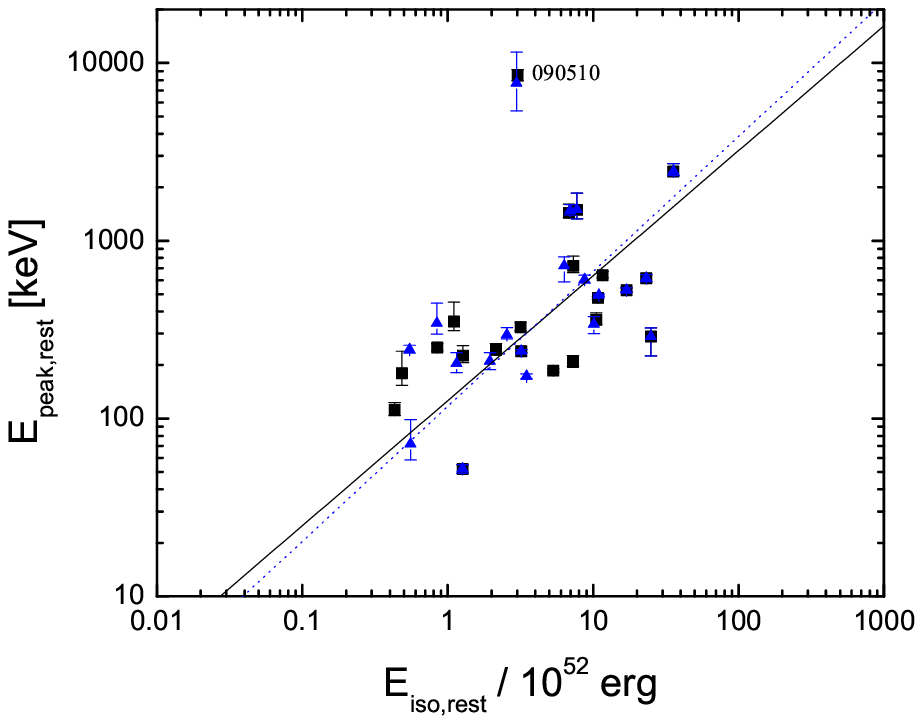}}
\caption{Alternative methods for calculating the relationship between E$_{\rm iso,rest}$-E$_{\rm peak,rest}$. In both panels, the black (square) icons and the solid black line are identical to those presented in Figure 8.  The dashed line represents the best linear bisector fit to the second data set for reference.  Short burst 090510 is labelled but not included in the calculations.  (a) Rest frame fluxes are calculated assuming a CPL spectrum (red/circle) instead of a Band function (black/square, identical to Fig 8).  This systematically underestimates the isotropic energy by a mean factor of about 1.4.  (b) Rest frame fluxes calculated using Band function parameters, even if a CPL gives a better fit (i.e. lower $\chi^2$; blue/triangle).  If $\beta$ cannot be constrained a typical value of -2.3 is assumed.  This gives a similar distribution to that produced in Figure 8 (black/square). }
\label{EpEiso2}
\end{figure*}

A sub-sample of 25 GRBs in GBM-BAT joint sample have measured redshifts.  With this value we can calculate the isotropic energy released which, combined with the rest-frame peak spectral energy, $E_{\rm peak,rest}$, can be used to test the empirical Amati relation \cite{amati02,amati03}.  Since our spectral fits are derived from time-integrated spectra, we can derive the mean energy flux $F$ during the extracted time scale $\Delta t$ (which is essentially $T_{90}$) in a desired energy band, which we uniformly adopt as the rest-frame 1-10$^4$ keV energy flux. We worked on only bursts whose time integrated spectra are fit by a Band function of CPL, therefore three bursts that are fit with a simple PL in the $z$-known sub-sample are excluded.  We then derived the broad-band isotropic $\gamma$-ray energy $E_{\rm iso} = 4\pi d_L^2 F \Delta t / (1+z)$, where $\Delta t$ is the time interval over which the spectral fit was derived, $d_L$ is the luminosity distance, and the factor of $(1+z)$ is the correction for cosmological time dilation.  If the spectrum is best described by a CPL then the best-fit pre-break slope ($\alpha$) is used with the assumption of a typical $\beta$ of -2.3 to construct a Band function.  Otherwise, the best-fit Band function parameters are used.  In Figure \ref{EpEiso2} we present the differences between deriving the flux from a CPL as opposed to a Band function.  The calculated fluence (and therefore E$_{\rm iso}$) is systematically underestimated by a mean factor of 1.4 and a maximum of 2 for this sample.  For the bursts that are best fit with a CPL we also refit the spectra with a Band function and use those spectral parameters, even though this model has a larger $\chi^2$, to derive $E_{\rm iso}$.  These results are also included in Figure \ref{EpEiso2} and show that the bursts tend to follow a similar, although not identical, distribution as the CPL fits.

\begin{figure*}
\centerline{\includegraphics[width=15cm]{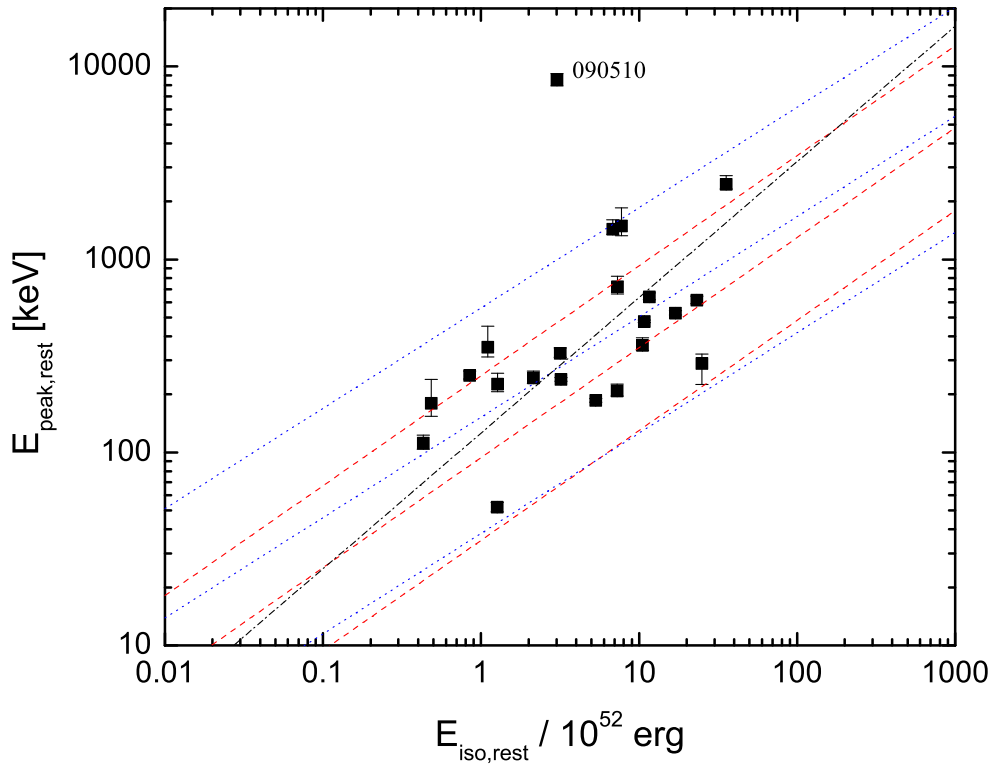}}
\caption{Relationship between the isotropic equivalent energy release and the peak energy of the time-integrated spectrum of the joint sample bursts.  Isotropic energies are derived using a Band function spectrum.  If the spectrum is best described by a CPL then the best-fit pre-break slope ($\alpha$) is used with the assumption of a typical $\beta$ of -2.3 to construct a Band function.  Superimposed are the best ordinary least squares bisector fit of the joint sample (black/dash-dot, slope=0.76) and the approximate best fit lines and 2$\sigma$ regions reported in Amati et al. (2008; red/dashed; slope=0.57) and Gruber et al. (2011; blue/solid; slope=0.52).  The shift to steeper slopes is produced from a lack of bursts above $E_{\rm iso} \sim 10^{54}$ erg, which is attributed to a selection effect of the sample. These bursts are the highest energy bursts detected and are \fermi LAT triggers and therefore do not show up in our sample of joint GBM-BAT bursts.  If these highest energy bursts are introduced to create a more complete sample, we show consistency with both slope values presented in the literature.  We also confirm the larger scatter and slightly harder spectra presented by Gruber et al. (2011).  Differences in the absolute values of $E_{\rm peak}$ are likely due to the derivation of the parameter using the joint BAT-GBM data.  The obvious outlier in the distribution is GRB 090510 (labelled), the only short burst ($T_{90}<2$ sec) in the redshift-known subsample.  This burst is not used in calculating the best-fit line.  Previous works (Amati et al. 2008; Zhang et al. 2009; Ghirlanda et al. 2009; Amati et al. 2008) have shown that these bursts follow a different relationship in the $E_{\rm peak,rest}-E_{\rm iso}$ plane and are not considered in the fitting of the slope of the relation.}
\label{EpEiso}
\end{figure*}

We fit the Amati relation sample with two different methods: ordinary least squares and ordinary least squares bisector fit, the latter an accepted method to fit linear models while taking into account some of the scatter of the data set.  We provide both fits for comparison, although the bisector method is likely a more accurate depiction of the true relation.  For the joint sample these methods give differing solutions that we explore below.  First, using the ordinary least squares method, our results are generally consistent with slopes for the $E_{\rm peak,rest} - E_{\rm iso}$ relation presented in the literature \cite{amati08,zhang09,ghirlanda09,ghirlanda10,amati10,gruber11}, having a slope of 0.5 but a larger normalization (see Figure \ref{EpEiso}).  Using the bisector fit we find that the slope steepens significantly to 0.76, with a lower intercept.  This steepening, however, is due to a selection effect in the data owing to a lack of bursts above an energy of about $10^{54}$ erg.  The highest energy bursts reported by \fermi (e.g. 080916C, 090323) are not simultaneously detected by \swift and are not included in our sample.  Adding in the three highest energy \fermi bursts by hand to simulate a more unbiased sample (data from D. Gruber, private communication), the relation drops down to 0.57, which is fully consistent with previous estimates and serves as a consistency check for our sample. In general, we derive a larger scatter in the Amati relation than the sample presented in works such as Amati et al. (2008, 2010). The results are consistent, however, with another independent analysis of the GBM bursts (Gruber et al. 2011).  The mean and median of the $E_{\rm iso}$ distribution of our sample are 8.4$\times 10^{52}$ erg and $6.07\times 10^{52}$ erg, respectively.  Figure \ref{EpEiso} shows the $E_{\rm peak,rest}-E_{\rm iso}$ relationship for the 20 bursts of the joint sample with known redshift with the linear bisector fit as well as an approximation of the fits by Gruber et al. (2011; blue/solid) and Amati et al. (2008; red/dashed) with their respective 2$\sigma$ regions.  The outlier of the distribution, having an $E_{\rm peak,rest}$ of about 8000 keV, is GRB 090510.  It is the only redshift-known short burst in the sample.  Previous works have shown that short ($T_{90}<2$ sec) bursts follow a different relationship and are not included in the slope determination of the longer bursts \cite{amati08,zhang09,ghirlanda09,amati10}.

\section{Comparison of Burst Durations in the BAT and GBM Bands}

We calculate the $T_{90}$ values observed in the GBM band for the bursts in our sample by using a Bayesian Block  method \cite{scargle98,richardson96}. The lightcurves we use are in 64 ms bins and are extracted from the brightest NaI detector. With the Bayesian Block  method we derive $T_{5}$ and $T_{95}$, the epochs when 5\% and 95\% of the total fluence were registered, respectively. We then derive $T_{90}$ by $T_{90}=T_{95}-T_{5}$ with the errors of $T_{5}$ and $T_{95}$ estimated with the bootstrap method and assuming that the observed errors in the lightcurves are log-normal. We generate 1000 mimic lightcurves from the error distributions and calculate their $T_{90}$ for a given burst. We make a Gaussian fit to the $T_{90}$ distributions then derive the value of $T_{90}$ and its 1$\sigma$ error. $T_{90}$ in the BAT band is taken from the Second BAT Catalog \cite{sakamoto11} (bursts before 2010) and GCN reports (bursts after 2010; see Table 3 for specific references).

\begin{figure*}
\centerline{\includegraphics[width=21cm]{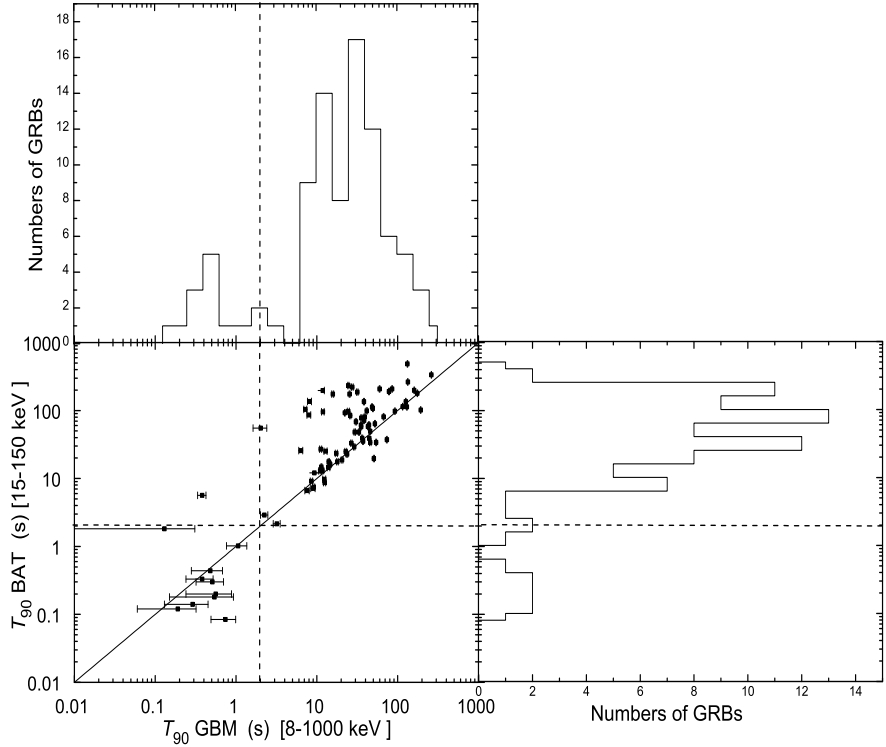}}
\caption{$T_{90}$ measures for the joint sample bursts as derived with the BAT (15-150 keV) and GBM (8-1000 keV)data and corresponding distributions.  The diagonal solid line is the line of slope = 1, while the dashed lines indicate a duration of 2 sec. Note the typically longer durations in the \swift band for the long and generally softer bursts while the trend reverses for the shorter, typically softer, bursts.}
\label{t901}
\end{figure*}

A comparison of the $T_{90}$ in the BAT and GBM bands along with their distributions is shown in  Figure \ref{t901}.  It is found that the distributions are still bimodal. The $T_{90}$ measured in the BAT band for long GRBs is usually larger than that in the GBM band. However, the $T_{90}$ measured in the BAT band for short GRBs is even shorter than that in the GBM band. Since BAT is more sensitive to softer emission than GBM, this fact is consistent with the soft/hard nature of long/short GRBs, respectively. In particular, fainter, softer emission can be picked up by BAT but not by GBM, so that $T_{90}$(BAT) can be much longer than $T_{90}$(GBM) in some bursts. Conversely, since short GRBs are typically hard, BAT may not be sensitive enough to pick up some emission episodes that GBM can. As a result, $T_{90}$ of short GRBs can be longer in the GBM band than in the BAT band. 

The difference in $T_{90}$ for different detectors may also affect the derivation of the isotropic energy. This is because some long bursts having shorter $T_{90}$ in the GBM band than BAT band may be sampling only the hardest and brightest part of the spectrum. For the small sample of six bursts above the Amati et al. $2\sigma$ region in Fig. \ref{EpEiso} we find that all have longer durations in the BAT bands with half of them significantly longer (difference $> 30\%$), which may indicate the existence of this spectrum sampling effect. This suggests that in order to fully characterize the durations, and hence, the physical origin of the bursts, multi-band data and a careful energy-dependent temporal analysis are essential (see Qin et al. 2012).

\section{Conclusions and Discussion}

Using the data available from \fermi and \swift we performed an analysis of the jointly detected GRB sample. The spectral analysis indicates that fitting the joint spectra shows consistency with values reported in the literature based solely on GBM data  \cite{nava11,bissaldi11}, and is able to reproduce a variety of empirical relations presented in the literature, including the $\Gamma^{\rm BAT}-E_{\rm peak}$ and Amati relations.  The joint fits do not, in general, give substantially different values for the spectral parameters from the GBM-only analyses.  

The updated sample of peak energies shows agreement with the relation between the PL slope of the BAT spectrum and $E_{\rm peak}$ derived previously by Sakamoto et al. (2009) and Zhang et al. (2007b) and indicates the relation's ability to estimate the value of the observed $E_{\rm peak}$ from a fit to the rather narrow \swift band.  The relation shows a fair amount of scatter, however, and we caution against using it for robust measurements of $E_{\rm peak}$, especially when compared to spectra from \fermi or other missions that often contain the spectral peak within the energy range of the detectors.  It may be suited, for example, to give a rough estimate of the values of $E_{\rm peak}$ for \swift bursts that have no additional observations. 

The redshift known subsample of bursts reproduces the general relationship in the $E_{\rm iso}-E_{\rm peak,rest}$ plane and further confirms the conclusions of Gruber et al. (2011) that show generally harder spectra and a larger scatter around the relation than previously believed.  Li (2007) proposed the argument that the $E_{\rm iso}-E_{\rm peak,rest}$ relation might evolve with redshift.  We believe that the sample of bursts with known redshift, especially the subsample of joint bursts, is still too small to test such a claim.  The necessary binning to test the hypothesis would thin out the sample sufficiently to make firm conclusions tenuous at best.  

Finally, the interesting discrepancy between the $T_{90}$ derived in the GBM and BAT detectors confirms the soft/hard nature of long/short GRBs, respectively, and suggests the need for a detailed energy-dependent temporal analysis of GRBs.

\section*{Acknowledgments}

We would like to thank Taka Sakamoto, Valerie Connaughton, Sylvia Zhu and Binbin Zhang for helpful discussions, David Gruber for providing his $E_{\rm iso}$ and $E_{\rm peak}$ data for comparison with our fits, and the anonymous referee for comments and suggestions that helped improve our manuscript. BZ and FJV acknowledge support from NASA (NNX10AD48G) and NSF (AST-0908362). EWL and YQ acknowledge the support by the National Natural Science Foundation of China (Grants 11025313, 10873002), National Basic Research Program (``973" Program) of China (Grant 2009CB824800), and Guangxi Science Foundation (2010GXNSFC013011, 2011-135).

\onecolumn
\begin{deluxetable}{ccccc}
\tabletypesize{\small}
\tablecaption{Joint sample bursts whose time-integrated spectra are best fit by a simple powerlaw model. $\Gamma$ is the slope of the powerlaw and $\Delta t$ the time interval used for the fit.  Spectra were fit using data in the energy range of 15-150 keV(BAT)+8 keV-40 MeV (GBM).  Reported errors are to the 90\% level.}
\tablewidth{0pt}
\tablehead{
\colhead{GRB} & 
\colhead{GBM ID} & 
\colhead{$\Delta t$} & 
\colhead{$\Gamma$} & 
\colhead{$\chi^2/dof$} \\
\colhead{name} &
\colhead{} &
\colhead{(s)} &
\colhead{} &
\colhead{} 
}
\startdata
080905B & 080905705 & 26.624 & 1.62$^{0.09}_{-0.09}$ & 526/530 \\
080928A & 080928628 & 18.432 & 1.87$^{0.07}_{-0.07}$ & 537/529 \\
090422A & 090422150 & 13.312 & 2.12$^{0.24}_{-0.21}$ & 527/532 \\
090518A & 090518080 & 7.168 & 1.65$^{0.08}_{-0.07}$ & 402/421 \\
090927A & 090927422 & 9.216 & 1.93$^{0.2}_{-0.17}$ & 689/587 \\
100619A & 100619015 & 102.404 & 1.83$^{0.05}_{-0.05}$ & 350/413 \\
100727A & 100727238 & 18.432 & 1.91$^{0.09}_{-0.09}$ & 545/533 \\
\enddata
\end{deluxetable}

\break

\begin{deluxetable}{cccccccc}
\tabletypesize{\small}
\tablecaption{Bursts whose time-integrated spectra are best fit by either a cutoff powerlaw or Band function model. Spectra were fit using data in the 15-150 keV (BAT)+8 keV-40 MeV (GBM) energy range. Errors are to the 90\% level.}
\tablewidth{0pt}
\tablehead{
\colhead{GRB} & 
\colhead{GBM ID} &
\colhead{Spectral} &
\colhead{$\Delta t^b$} & 
\colhead{$\alpha$} & 
\colhead{$\beta$} & 
\colhead{$E_{\rm peak}$} & 
\colhead{$\chi^2/dof$} \\
\colhead{name} &
\colhead{} &
\colhead{Model$^a$} &
\colhead{(s)} &
\colhead{} &
\colhead{} &
\colhead{(keV)} &
\colhead{} 
}
\startdata
080714A & 080714745 & CPL & 28.672 & 1.22$^{0.13}_{-0.15}$ & -- & 155$^{54}_{-31}$ & 401/395 \\
080725A & 080725435 & CPL & 36.864 & 1.12$^{0.07}_{-0.08}$ & -- & 309$^{85}_{-57}$ & 456/408 \\
080727C & 080727964 & CPL & 77.825 & 1.13$^{0.08}_{-0.09}$ & -- & 191$^{41}_{-28}$ & 711/526 \\
080804A & 080804972 & CPL & 26.62 & 0.56$^{0.12}_{-0.13}$ & -- & 200$^{30}_{-23}$ & 465/408 \\
080810A & 080810549 & CPL & 69.633 & 1.23$^{0.06}_{-0.07}$ & -- & 564$^{269}_{-140}$ & 581/528 \\
080905A & 080905499 & CPL & 1.152 & 0.25$^{0.37}_{-0.5}$ & -- & 586$^{431}_{-182}$ & 283/267 \\
080916A & 080916406 & CPL & 69.633 & 1.24$^{0.1}_{-0.11}$ & -- & 134$^{31}_{-19}$ & 402/410 \\
081008A & 081008832 & CPL & 60.417 & 1.19$^{0.1}_{-0.11}$ & -- & 503$^{359}_{-167}$ & 591/526 \\
081012 A& 081012549 & CPL & 18.432 & 0.05$^{0.18}_{-0.38}$ & -- & 256$^{71}_{-46}$ & 506/526 \\
081024A & 081024245 & CPL & 0.256 & 1.05$^{0.24}_{-0.33}$ & -- & 2170$^{2074}_{-2074}$ & 104/103 \\
081102A & 081102739 & CPL & 50.177 & 0.84$^{0.25}_{-0.28}$ & -- & 80$^{16}_{-10}$ & 570/529 \\
081109A & 081109293 & CPL & 40.961 & 1.44$^{0.14}_{-0.15}$ & -- & 133$^{68}_{-31}$ & 548/530 \\
081121A & 081121858 & Band & 21.501 & 0.53$^{0.16}_{-0.18}$ & 2.16$^{0.19}_{-0.14}$ & 175$^{31}_{-25}$ & 398/391 \\
081126A & 081126899 & CPL & 51.171 & 0.96$^{0.1}_{-0.12}$ & -- & 249$^{73}_{-47}$ & 425/410 \\
081221A & 081221681 & Band & 39.422 & 0.86$^{-0.06}_{0.06}$ & 3.14$^{0.6}_{-0.28}$ & 81$^{3}_{-3}$ & 568/410 \\
081222A & 081222204 & Band & 14.336 & 0.85$^{0.07}_{-0.08}$ & 2.39$^{0.33}_{-0.22}$ & 140$^{14}_{-13}$ & 460/371 \\
081226A & 081226044 & CPL & 0.384 & 0.9$^{0.28}_{-0.33}$ & -- & 418$^{496}_{-142}$ & 121/120 \\
090102A & 090102122 & CPL & 37.888 & 0.98$^{0.04}_{-0.04}$ & -- & 443$^{46}_{-39}$ & 499/410 \\
090113A & 090113778 & CPL & 12.288 & 1.29$^{0.19}_{-0.23}$ & -- & 138$^{100}_{-39}$ & 346/348 \\
090129A & 090129880 & CPL & 15.36 & 1.53$^{0.09}_{-0.1}$ & -- & 163$^{91}_{-40}$ & 343/376 \\
090423A & 090423330 & CPL & 14.336 & 0.87$^{0.37}_{-0.73}$ & -- & 52$^{8}_{-6}$ & 517/506 \\
090424A & 090424592 & Band & 19.71 & 0.95$^{0.02}_{-0.02}$ & 2.89$^{0.26}_{-0.17}$ & 155$^{5}_{-5}$ & 844/525 \\
090509A & 090509215 & CPL & 26.624 & 0.9$^{0.2}_{-0.23}$ & -- & 221$^{112}_{-58}$ & 572/527 \\
090510A & 090510016 & CPL & 0.896 & 0.84$^{0.05}_{-0.05}$ & -- & 4482$^{658}_{-581}$ & 329/304 \\
090519A & 090519881 & CPL & 18.432 & 0.51$^{0.28}_{-0.34}$ & -- & 296$^{171}_{-85}$ & 532/518 \\
090531B & 090531775 & CPL & 3.072 & 0.93$^{0.15}_{-0.16}$ & -- & 1911$^{1067}_{-713}$ & 371/347 \\
090621B & 090621922 & CPL & 0.128 & 0.47$^{0.3}_{-0.34}$ & -- & 577$^{826}_{-230}$ & 57.2/77 \\
090708A & 090708152 & CPL & 14.336 & 0.8$^{0.44}_{-0.51}$ & -- & 54$^{11}_{-8}$ & 460/496 \\
090709B & 090709630 & CPL & 20.48 & 0.92$^{0.2}_{-0.22}$ & -- & 121$^{32}_{-20}$ & 507/530 \\
090712A & 090712160 & CPL & 53.248 & 1.06$^{0.11}_{-0.12}$ & -- & 588$^{566}_{-204}$ & 492/529 \\
090813A & 090813174 & CPL & 9.725 & 1.51$^{0.13}_{-0.14}$ & -- & 100$^{31}_{-18}$ & 344/338 \\
090904B & 090904058 & CPL & 74.663 & 1.28$^{0.08}_{-0.08}$ & -- & 139$^{19}_{-14}$ & 554/529 \\
090912A & 090912660 & CPL & 123.902 & 0.8$^{0.22}_{-0.24}$ & -- & 72$^{9}_{-7}$ & 536/530 \\
090926B & 090926914 & CPL & 64.513 & 0.2$^{0.14}_{-0.15}$ & -- & 83$^{4}_{-4}$ & 520/529 \\
091020A & 091020900 & CPL & 28.672 & 1.36$^{0.08}_{-0.08}$ & -- & 266$^{97}_{-59}$ & 382/412 \\
091026A & 091026550 & CPL & 15.36 & 1.31$^{0.15}_{0.17}$ & -- & 272$^{235}_{-99}$ & 410/378 \\
091102A & 091102607 & CPL & 9.216 & 1.03$^{0.15}_{-0.17}$ & -- & 439$^{364}_{-157}$ & 514/521 \\
091127A & 091127976 & Band & 11.006 & 1.21$^{0.13}_{-0.15}$ & 2.25$^{0.04}_{-0.04}$ & 35$^{3}_{-3}$ & 466/366 \\
091208B & 091208410 & CPL & 13.312 & 1.36$^{0.09}_{-0.09}$ & -- & 118$^{22}_{-15}$ & 486/375 \\
091221A & 091221870 & CPL & 35.841 & 1.07$^{0.07}_{-0.07}$ & -- & 275$^{59}_{-41}$ & 721/529 \\
100111A & 100111176 & CPL & 16.374 & 1.59$^{0.17}_{-0.21}$ & -- & 203$^{122}_{-122}$ & 706/649 \\
100117A & 100117879 & CPL & 0.192 & 0.14$^{0.39}_{-0.52}$ & -- & 380$^{194}_{-122}$ & 93.1/86 \\
100206A & 100206563 & CPL & 0.256 & 0.64$^{0.2}_{-0.21}$ & -- & 763$^{865}_{-269}$ & 109/126 \\
100504A & 100504806 & CPL & 28.672 & 1.03$^{0.22}_{-0.24}$ & -- & 85$^{19}_{-11}$ & 582/528 \\
100522A & 100522157 & CPL & 41.985 & 1.54$^{0.22}_{-0.18}$ & -- & 59$^{21}_{-9}$ & 393/412 \\
100615A & 100615083 & Band & 40.574 & 0.91$^{0.29}_{-0.38}$ & 2.07$^{0.20}_{-0.09}$ & 50$^{15}_{-10}$ & 437/410 \\
100625A & 100625773 & CPL & 0.32 & 0.69$^{0.11}_{-0.12}$ & -- & 572$^{189}_{-131}$ & 160/143 \\
100704A & 100704149 & CPL & 23.552 & 0.91$^{0.09}_{-0.09}$ & -- & 187$^{33}_{-23}$ & 584/528 \\
100728A & 100728095 & CPL & 191.487 & 0.8$^{0.03}_{-0.03}$ & -- & 313$^{19}_{-17}$ & 538/411 \\
100728B & 100728439 & CPL & 10.237 & 0.93$^{0.18}_{-0.2}$ & -- & 105$^{22}_{-15}$ & 537/501 \\
100802A & 100802240 & CPL & 41.985 & 0.57$^{0.37}_{-0.41}$ & -- & 83$^{22}_{-12}$ & 684/650 \\
100814A & 100814160 & CPL & 158.722 & 1.06$^{0.11}_{-0.12}$ & -- & 147$^{33}_{-21}$ & 566/526 \\
100816A & 100816026 & CPL & 4.542 & 0.5$^{0.11}_{-0.12}$ & -- & 139$^{13}_{-11}$ & 437/364 \\
100906A & 100906576 & Band & 121.347 & 1.44$^{0.11}_{-0.59}$ & 1.99$^{0.23}_{-0.2}$ & 106$^{35}_{-64}$ & 589/526 \\
100924A & 100924165 & CPL & 8.061 & 1.16$^{0.14}_{-0.17}$ & -- & 172$^{68}_{-36}$ & 472/437 \\
101008A & 101008697 & CPL & 14.336 & 1.24$^{0.17}_{-0.22}$ & -- & 701$^{597}_{-597}$ & 422/410 \\
101011A & 101011707 & CPL & 27.645 & 1.04$^{0.2}_{-0.28}$ & -- & 389$^{999}_{-190}$ & 520/524 \\
101024A & 101024486 & CPL & 13.312 & 0.91$^{0.41}_{-0.45}$ & -- & 48$^{6}_{-5}$ & 543/504 \\
101213A & 101213451 & CPL & 47.105 & 1.26$^{0.1}_{-0.12}$ & -- & 423$^{410}_{-143}$ & 380/411 \\
101219B & 101219686 & CPL & 51.201 & 0.08$^{0.39}_{-0.45}$ & -- & 72$^{11}_{-8}$ & 738/650 \\
110102A & 110102788 & CPL & 148.486 & 1.46$^{0.06}_{-0.07}$ & -- & 479$^{695}_{-196}$ & 403/411 \\
110106B & 110106893 & CPL & 35.84 & 1.4$^{0.18}_{-0.2}$ & -- & 111$^{59}_{-26}$ & 509/528 \\
110119A & 110119931 & CPL & 68.609 & 1.15$^{0.11}_{-0.12}$ & -- & 224$^{110}_{-56}$ & 555/527 \\
110201A & 110201399 & CPL & 12.288 & 0.93$^{0.12}_{-0.15}$ & -- & 730$^{641}_{-311}$ & 612/595 \\
110213A & 110213220 & CPL & 43.009 & 1.56$^{0.09}_{-0.1}$ & -- & 85$^{18}_{-12}$ & 411/410 \\
110318A & 110318552 & Band & 20.48 & 0.82$^{0.16}_{-0.19}$ & 2.3$^{0.4}_{-0.18}$ & 82$^{13}_{-11}$ & 446/394 \\
110402A & 110402009 & CPL & 38.526 & 1.4$^{0.11}_{-0.13}$ & -- & 914$^{3248}_{-495}$ & 618/649 \\
110412A & 110412315 & CPL & 20.48 & 0.64$^{0.24}_{-0.26}$ & -- & 84$^{12}_{-9}$ & 532/528 \\
\enddata
\tablenotetext{a}{CPL = Cutoff powerlaw or powerlaw + exponential. 100 keV normalization energy.}
\tablenotetext{b}{Duration of time bin for the evaluation of the time-integrated spectra. This is determined by the duration of the signal of the GBM bursts in RMfit}
\tablenotetext{c}{Several bursts in the table (080727C, 081221, 081222, 090424, 091127, 091208B, 100728A) have $\chi^2$ values larger than about 1.3.  These bursts are usually long (20 s $> \Delta t >$ 200s) and/or have multi-peak structure, and show significant spectral evolution over the duration of the burst.  Doing a simple time-dependent analysis show the fits improve but the spectral evolution likely drives the larger residuals for the full time interval.}
\end{deluxetable}

\begin{deluxetable}{ccccc}
\tabletypesize{\small}
\tablecaption{GBM and BAT $T_{90}$ values in the 8-1000 keV and 15-150 keV energy ranges, respectively. GBM values are derived with the method described in \S 2.2 and BAT figures are referenced from the literature (see reference column for appropriate citations).}
\tablewidth{0pt}
\tablehead{
\colhead{GRB} & 
\colhead{GBM ID} & 
\colhead{GBM $T_{90}$} &
\colhead{BAT $T_{90}$} &
\colhead{Reference$^a$} \\
\colhead{name} &
\colhead{} &
\colhead{(s)} &
\colhead{(s)} &
\colhead{} 
}
\startdata
080714 & GRB080714745 & 6.34$\pm$0.36 & 25.81 & bat2 \\
080725 & GRB080725435 & 22.21$\pm$0.23 & 92.74 & bat2 \\
080727C & GRB080727964 & 35.55$\pm$0.59 & 77.61 & bat2 \\
080804 & GRB080804972 & 73.41$\pm$0.93 & 37.19 & bat2 \\
080810 & GRB080810549 & 49.34$\pm$0.63 & 107.67 & bat2 \\
080905A & GRB080905499 & 1.06$\pm$0.30 & 1.02 & bat2 \\
080905B & GRB080905705 & 192.8$\pm$1.15 & 101.62 & bat2 \\
080916A & GRB080916406 & 44.26$\pm$0.72 & 61.35 & bat2 \\
080928 & GRB080928628 & 24.54$\pm$0.44 & 233.66 & bat2 \\
081008 & GRB081008832 & 175.2$\pm$1.16 & 179.52 & bat2 \\
081012 & GRB081012549 & 12.8$\pm$0.56 & 25.2 & bat2 \\
081024A & GRB081024245 & 0.13$\pm$0.18 & 1.82 & bat2 \\
081025 & GRB081025349 & 23.62$\pm$0.49 & 22.78 & bat2 \\
081101 & GRB081101491 & 0.54$\pm$0.39 & 0.18 & bat2 \\
081102 & GRB081102739 & 29.47$\pm$0.59 & 48.19 & bat2 \\
081109A & GRB081109293 & 27.46$\pm$0.65 & 221.49 & bat2 \\
081121 & GRB081121858 & 17.98$\pm$0.42 & 17.67 & bat2 \\
081126 & GRB081126899 & 35.36$\pm$0.46 & 57.65 & bat2 \\
081221 & GRB081221681 & 45.82$\pm$0.99 & 33.91 & bat2 \\
081222 & GRB081222204 & 26.75$\pm$0.53 & 33 & bat2 \\
081226A & GRB081226044 & 0.48$\pm$0.20 & 0.44 & bat2 \\
090102 & GRB090102122 & 29.02$\pm$0.54 & 29.3 & bat2 \\
090113 & GRB090113778 & 8.58$\pm$0.29 & 9.1 & bat2 \\
090129 & GRB090129880 & 14.02$\pm$0.23 & 17.66 & bat2 \\
090422 & GRB090422150 & -- & 8.47 & bat2 \\
090423 & GRB090423330 & 12.35$\pm$0.50 & 9.77 & bat2 \\
090424 & GRB090424592 & 45.79$\pm$0.61 & 49.47 & bat2 \\
090509 & GRB090509215 & 261.18$\pm$1.08 & 336.38 & bat2 \\
090510 & GRB090510016 & 0.38$\pm$0.045 & 5.66 & bat2 \\
090516A & GRB090516353 & 85.28$\pm$0.54 & 208 & bat2 \\
090518 & GRB090518080 & 7.97$\pm$0.44 & 85.82 & bat2 \\
090519 & GRB090519881 & 42.88$\pm$0.57 & 58.26 & bat2 \\
090531B & GRB090531775 & 2.016$\pm$0.39 & 55 & bat2 \\
090618 & GRB090618353 & 130.24$\pm$1.05 & 113.34 & bat2 \\
090621B & GRB090621922 & 0.29$\pm$0.16 & 0.14 & bat2 \\
090708 & GRB090708152 & 12.48$\pm$0.39 & 8.7 & bat2 \\
090709B & GRB090709630 & 11.14$\pm$0.32 & 27.02 & bat2 \\
090712 & GRB090712160 & 31.68$\pm$0.62 & 186.68 & bat2 \\
090813 & GRB090813174 & 8.96$\pm$0.55 & 7.14 & bat2 \\
090904B & GRB090904058 & 52.35$\pm$0.59 & 64 & bat2 \\
090912 & GRB090912660 & 126.37$\pm$0.54 & 135.52 & bat2 \\
090926B & GRB090926914 & 41.34$\pm$0.62 & 99.28 & bat2 \\
090927 & GRB090927422 & 3.2$\pm$0.32 & 2.16 & bat2 \\
091020 & GRB091020900 & 44.67$\pm$0.65 & 38.92 & bat2 \\
091024 & GRB091024372 & 48.26$\pm$0.62 & 112.28 & bat2 \\
091026 & GRB091026550 & 15.74$\pm$0.46 & 174 & bat2 \\
091102 & GRB091102607 & 7.58$\pm$0.46 & 6.65 & bat2 \\
091112 & GRB091112737 & 50.85$\pm$0.82 & 19.59 & bat2 \\
091127 & GRB091127976 & 9.15$\pm$0.26 & 7.42 & bat2 \\
091208B & GRB091208410 & 11.39$\pm$0.14 & 14.8 & bat2 \\
091221 & GRB091221870 & 30.432$\pm$0.62 & 68.49 & bat2 \\
100111A & GRB100111176 & 11.01$\pm$0.46 & 12.9 & 10322 \\
100117A & GRB100117879 & 0.51$\pm$0.19 & 0.3 & 10338 \\
100206A & GRB100206563 & 0.19$\pm$0.13 & 0.12 & 10379 \\
100212 & GRB100212588 & 8.10$\pm$0.46 & 136 & 10404 \\
100413A & GRB100413732 & 78.11$\pm$1.07 & 191 & 10600 \\
100427A & GRB100427356 & 10.43$\pm$0.41 & -- & -- \\
100504A & GRB100504806 & 24.51$\pm$0.73 & 97.3 & 10716 \\
100522A & GRB100522157 & 37.38$\pm$0.59 & 35.3 & 10788 \\
100615A & GRB100615083 & 36.42$\pm$0.59 & 39 & 10850 \\
100619A & GRB100619015 & 92.61$\pm$0.34 & 97.5 & 10864 \\
100625A & GRB100625773 & 0.38$\pm$0.14 & 0.33 & 10891 \\
100704A & GRB100704149 & 11.81$\pm$0.54 & 197.5 & 10932 \\
100727A & GRB100727238 & 25.92$\pm$0.47 & 84 & 11001 \\
100728A & GRB100728095 & 159.97$\pm$0.76 & 198.5 & 11018 \\
100728B & GRB100728439 & 9.34$\pm$0.46 & 12.1 & 11023 \\
100802A & GRB100802240 & 132.26$\pm$0.84 & 487 & 11035 \\
100814A & GRB100814160 & 25.41$\pm$0.29 & 174.5 & 11094 \\
100816A & GRB100816026 & 2.24$\pm$0.23 & 2.9 & 11111 \\
100906A & GRB100906576 & 115.56$\pm$0.68 & 114.4 & 11234 \\
100924A & GRB100924165 & 11.87$\pm$0.43 & 96 & 11296 \\
101008A & GRB101008697 & 7.2$\pm$0.30 & 104 & 11327 \\
101011A & GRB101011707 & 38.50$\pm$0.62 & 71.5 & 11332 \\
101023A & GRB101023951 & 66.94$\pm$0.55 & 80.8 & 11367 \\
101024A & GRB101024486 & 20.51$\pm$0.54 & 18.7 & 11374 \\
101129A & GRB101129652 & 0.544$\pm$0.17 & -- & -- \\
101201A & GRB101201418 & 83.62$\pm$0.68 & -- & -- \\
101213A & GRB101213451 & 38.24$\pm$0.56 & 135 & 11453 \\
101219B & GRB101219686 & 54.21$\pm$0.75 & 34 & 11475 \\
101224A & GRB101224227 & 0.56$\pm$0.32 & 0.2 & 11486 \\
110102A & GRB110102788 & 133.76$\pm$0.34 & 264 & 11511 \\
110106B & GRB110106893 & 22.78$\pm$0.62 & 24.8 & 11533 \\
110119A & GRB110119931 & 59.87$\pm$0.59 & 208 & 11584 \\
110128A & GRB110128073 & -- & 30.7 & 11614 \\
110201A & GRB110201399 & 11.712$\pm$0.51 & 13 & 11624 \\
110207A & GRB110207470 & 39.01$\pm$0.52 & 80.3 & 11664 \\
110213A & GRB110213220 & 33.12$\pm$0.46 & 48 & 11714 \\
110318A & GRB110318552 & 14.66$\pm$0.41 & 16 & 11802 \\
110319B & GRB110319815 & 13.95$\pm$0.52 & 14.5 & 11818 \\
110402A & GRB110402009 & 35.19$\pm$0.53 & 60.9 & 11866 \\
110412A & GRB110412315 & 17.4$\pm$0.53 & 23.4 & 11929 \\
110420B & GRB110420946 & 0.74$\pm$0.25 & 0.084 & 11946 \\
\enddata
\tablenotetext{a}{References for BAT $T_{90}$: bat2 - Second BAT Catalog \cite{sakamoto11}. Other numbers in this column correspond to GCN reports.  See bibliography for full record information.}
\end{deluxetable}

\label{lastpage}

\begin{thebibliography}{99}

\bibitem[Amati et al. 2002]{amati02} Amati L., et al. 2002, A\&A, 390, 81
\bibitem[Amati 2003]{amati03} Amati L. 2003, ChJAA Suppl., 3 455 
\bibitem[Amati et al. 2008]{amati08} Amati L., Guidorzi C., Frontera F., Della Valle M., Finelli F., Landi R., Montanari E. 2008, MNRAS, 391, 577
\bibitem[Amati 2010]{amati10} Amati L. 2010, arXiv:1002.2232
\bibitem[Band \& Preece 2005]{band05} Band D. L., Preece R. D. 2005, ApJ, 627, 319
\bibitem[Band et al. 1993]{band93} Band D. L., et al. 1993, ApJ, 413, 281
\bibitem[]{} Barthelmy S., Sakamoto T., Holland S. 2010, GCN, 10600, 1
\bibitem[]{} Barthelmy S., et al. 2010a, GCN, 10788, 1
\bibitem[]{} Barthelmy S., et al. 2010b, GCN, 10891, 1
\bibitem[]{} Barthelmy S., et al. 2010c, GCN, 11023, 1
\bibitem[]{} Barthelmy S., et al. 2010d, GCN, 11234, 1
\bibitem[]{} Barthelmy S., et al. 2010e, GCN, 11296, 1
\bibitem[]{} Barthelmy S., et al. 2010f, GCN, 11327, 1
\bibitem[]{} Baumgartner W., et al. 2010a, GCN, 11035, 1
\bibitem[]{} Baumgartner W., et al. 2010b, GCN, 11584, 1
\bibitem[]{} Baumgartner W., et al. 2010c, GCN, 11929, 1
\bibitem[]{} Barthelmy S., et al. 2010a, GCN, 11714, 1
\bibitem[]{} Barthelmy S., et al. 2010b, GCN, 11802, 1
\bibitem[Bissaldi et al. 2011]{bissaldi11} Bissaldi E., et al. 2011, ApJ, 733, 97
\bibitem[]{} Cummings J., et al. 2010a, GCN, 10932, 1
\bibitem[]{} Cummings J., et al. 2010b, GCN, 11453, 1
\bibitem[]{} Cummings J., et al. 2010c, GCN, 11475, 1
\bibitem[]{} Cummings J., et al. 2010d, GCN, 11614, 1
\bibitem[]{} Cui X. H., Liang E. W., Lu R. J. 2005, ChJAA, 5, 151 
\bibitem[Gehrels et al. 2004]{gehrels04} Gehrels N. et al. 2004, ApJ, 611, 1005
\bibitem[Ghirlanda et al. 2004]{ghirlanda04} Ghirlanda G., Ghisellini, G., Lazzati, D. 2004, ApJ, 616, 331
\bibitem[Ghirlanda et al. 2009]{ghirlanda09} Ghirlanda G., Nava L., Ghisellini G., Celotti A., Firmani C. 2009, A\&A, 496, 585
\bibitem[Ghirlanda et al. 2010]{ghirlanda10} Ghirlanda G., Nava L., Ghisellini G. 2010, A\&A, 511, A43
\bibitem[Gruber et al. 2011]{gruber11} Gruber D., et al. 2011, A\&A, 531, A20
\bibitem[Kocevski 2011]{kocevski11} Kocevski D. 2012, ApJ, 747, 146
\bibitem[Kouveliotou et al. 1993]{kouveliotou93} Kouveliotou, C. et al. 1993, ApJ, 413, 101
\bibitem[Liang \& Zhang 2005]{liangzhang05} Liang E. W., Zhang B. 2005, ApJ, 633, 611
\bibitem[Li 2007]{li07} Li L.-X. 2007, MNRAS, 379, 55L
\bibitem[]{} Krimm H., et al. 2010a, GCN, 10322, 1
\bibitem[]{} Krimm H., et al. 2010b, GCN, 11094, 1
\bibitem[]{} Krimm H., et al. 2010c, GCN, 11624, 1
\bibitem[]{} Markwardt C., et al. 2010a, GCN, 10338, 1
\bibitem[]{} Markwardt C., et al. 2010b, GCN, 11111, 1
\bibitem[]{} Markwardt C., et al. 2010c, GCN, 11332, 1
\bibitem[]{} Markwardt C., et al. 2010d, GCN, 11486, 1
\bibitem[]{} Markwardt C., et al. 2010e, GCN, 11946, 1
\bibitem[Nakar \& Piran 2005]{nakar05} Nakar E., Piran T. 2005, MNRAS, 360, L73
\bibitem[Nava et al. 2011]{nava11} Nava L., Ghirlanda G., Ghisellini G., Celotti A. 2011, A\&A, 530, A21
\bibitem[]{} Palmer D., et al. 2010a, GCN, 10716, 1
\bibitem[]{} Palmer D., et al. 2010b, GCN, 10850, 1
\bibitem[]{} Palmer D., et al. 2010c, GCN, 11664, 1
\bibitem[]{} Palmer D., et al. 2010d, GCN, 11818, 1
\bibitem[Preece et al. 2000]{preece00} Preece R. D., Briggs M. S., Mallozzi R. S., Paciesas W. S., Band D. L. 2000, ApJS, 126, 19
\bibitem[Qin et al. 2012]{qin12} Qin, Y. et al. 2012, arXiv:1205.1188
\bibitem[Richardson et al. 1996]{richardson96} Richardson G., Koshut T., Paciesas W., Kouveliotou C. 1996, AIPC, 384, 87
\bibitem[Sakamoto et al. 2011]{sakamoto11} Sakamoto T., et al. 2011, PASJ, 63, 215 
\bibitem[]{} Sakamoto T., et al. 2010a, GCN, 10379, 1
\bibitem[]{} Sakamoto T., et al. 2010b, GCN, 11511, 1
\bibitem[Sakamoto et al. 2009]{sakamoto09} Sakamoto T., et al. 2009, ApJ, 693, 922
\bibitem[Sakamoto et al. 2006]{sakamoto06} Sakamoto T., et al. 2006, ApJ, 636, L73
\bibitem[Sakamoto et al. 2004]{sakamoto04} Sakamoto T., et al. 2004, ApJ, 602, 875
\bibitem[Scargle 1998]{scargle98} Scargle J. D., 1998, ApJ, 504, 405
\bibitem[]{} Stamatikos M., et al. 2010a, GCN, 10864, 1
\bibitem[]{} Stamatikos M., et al. 2010b, GCN, 11001, 1
\bibitem[]{} Stamatikos M., et al. 2010c, GCN, 11367, 1
\bibitem[]{} Stamatikos M., et al. 2010d, GCN, 11866, 1
\bibitem[]{} Ukwatta T., et al. 2010a, GCN, 10404, 1
\bibitem[]{} Ukwatta T., et al. 2010b, GCN, 11018, 1
\bibitem[]{} Ukwatta T., et al. 2010c, GCN, 11374, 1
\bibitem[]{} Ukwatta T., et al. 2010d, GCN, 11533, 1
\bibitem[Yonetoku et al. 2004]{yonetoku04} Yonetoku D., et al. 2004, ApJ, 571, 876
\bibitem[Zhang et al. 2007]{zhang07a} Zhang B., et al. 2007a, ApJ, 655, 989
\bibitem[Zhang et al. 2007]{zhang07b} Zhang B., et al. 2007b, ApJ, 655, L25
\bibitem[Zhang et al. 2009]{zhang09} Zhang B., et al. 2009, ApJ, 703, 1696
\bibitem[Zhang et al. 2011]{zhang11} Zhang B. B., et al. 2011, ApJ, 730, 141

\end{thebibliography}
\end{document}